\documentclass[a4paper]{article}

\usepackage{booktabs} \usepackage{cite}

\usepackage{amsmath,amssymb,amsfonts}
\usepackage{algorithmic}
\usepackage{graphicx}
\usepackage{subfigure}
\usepackage{textcomp}
\usepackage{siunitx}
\usepackage{xcolor}
\usepackage{xspace}
\usepackage{comment}
\usepackage{tikz}
\usepackage{tikz-qtree}

\usepackage[colorlinks,urlcolor=blue,linkcolor=magenta,citecolor=red,linktocpage=true]{hyperref}

\newcommand{\FOSS}{FOSS\xspace}
\newcommand{\RQi}{{\bf RQ1}\xspace}
\newcommand{\RQii}{{\bf RQ2}\xspace}

\newcommand{\SWH}{Software Heritage\xspace}

\newcommand{\errel}[1]{\textsc{\small #1}}

\newtheorem{definition}{Definition}

\def\figstretch{0.7}

\usepackage{a4wide}

\begin{document}
\title{Growth and Duplication of Public Source Code over Time: Provenance
  Tracking at Scale}

\newcommand{\email}[1]{\href{mailto:#1}{\tt #1}}

\def\affgrouss{  University Paris Diderot, France,
  \email{guillaume.rousseau@univ-paris-diderot.fr}
}
\def\affrdc{  Inria and University Paris Diderot, France,
  \email{roberto@dicosmo.org}
}
\def\affzack{  University Paris Diderot and Inria, France,
  \email{zack@irif.fr}
}

\author{  Guillaume Rousseau    \footnote{\affgrouss}
  \and
  Roberto Di~Cosmo    \footnote{\affrdc}
  \and
  Stefano Zacchiroli    \footnote{\affzack}
}

\maketitle

\begin{abstract}

  We study the evolution of the largest known corpus of publicly available
  source code, i.e., the Software Heritage archive (4B unique source code
  files, 1B commits capturing their development histories across 50M software
  projects). On such corpus we quantify the growth rate of original,
  never-seen-before source code files and commits. We find the growth rates to
  be exponential over a period of more than 40 years.

  We then estimate the multiplication factor, i.e., how much the same artifacts
  (e.g., files or commits) appear in different contexts (e.g., commits or
  source code distribution places). We observe a combinatorial explosion in the
  multiplication of identical source code files across different commits.

  We discuss the implication of these findings for the problem of tracking the
  provenance of source code artifacts (e.g., where and when a given source code
  file or commit has been observed in the wild) for the entire body of publicly
  available source code.  To that end we benchmark different data models for
  capturing software provenance information at this scale and growth rate.  We
  identify a viable solution that is deployable on commodity hardware and
  appears to be maintainable for the foreseeable future.

\end{abstract}

\section{Introduction} \label{sec:intro}

Over the last three decades, software development has been revolutionized under
the combined effect of the massive adoption of free and open source software
(\FOSS), and the popularization of collaborative development platforms like
GitHub, Bitbucket, and SourceForge~\cite{Squire17}, which have sensibly reduced
the cost of collaborative software development and offered a place where
historical software can be stored~\cite{SpinellisUnix2017}. One important
consequence of this revolution is that the source code and development history
of tens of millions of software projects are nowadays public, making an
unprecedented corpus available to software evolution scholars. We will refer to
this corpus as \emph{public source code} in this paper.

Many research studies have been conducted \emph{on subsets} of all public
source code, looking for patterns of interest for software engineering, ranging
from the study of code clones~\cite{SvajlenkoR17, SemuraYCI17,
  ThummalapentaCAP10} to automated vulnerability detection and
repair~\cite{Li2017,Grieco2016, MartinezM15}, from code
recommenders~\cite{Zeller2007, ZimmermannWDZ04} to software licence analysis
and compliance~\cite{GermanLicense17, VendomeLicence2015}.

Scaling up similar studies to the whole corpus, and making them reproducible,
is a significant challenge. In the absence of a common infrastructure providing
a \emph{reference archive} of pubic source code development, scholars have used
popular development platforms like GitHub as surrogates. But development
platforms are not archives: projects on GitHub come and go,\footnote{For
  example, hundreds of thousands of projects migrated from GitHub to GitLab.com
  in the days following the acquisition of GitHub by Microsoft in summer 2018,
  see~\url{https://about.gitlab.com/2018/06/03/movingtogitlab/}.} making
reproducibility a moving target. And while GitHub is the most popular
development platform today, millions of projects are developed elsewhere,
including very high profile ones like GNOME.\footnote{See
  \url{https://www.gnome.org/news/2018/05/gnome-moves-to-gitlab-2/}}

\SWH~\cite{swh-ipres-2017, swh-cacm-2018}---with its mission to collect,
preserve, and make accessible all public source code together with its
development history---offers an opportunity to change this state of affairs.
The project has amassed the largest public source code corpus to date, with
more than 80 millions software projects archived from GitHub, GitLab, PyPI, and
Debian, growing by the day.

In this paper we leverage \SWH to perform the first study on the evolution of
public source code. First, we look into the production of \emph{original}
source code artifacts over time, that is, the amount of source code files or
commits that have never been published before (e.g., in other VCS repositories
or distributed tarballs/packages) across the entire corpus. Our first research
question is:
\begin{description}

\item[\RQi] how does the public production of \emph{original}, i.e., never
  published before, source code artifacts, and in particular files and commits,
  evolve over time? what are the respective growth rates?

\end{description}

To answer this we perform an extensive study of the \SWH archive, continuing a
long tradition of software evolution studies~\cite{SurveyCrowston2008,
  LawsEvolutionHerraizRRG13, debsources-ese-2016, HattonSG17,MLgitarchive18},
which we extend by several orders of magnitude and perform over a period of
more than 40 years. We show evidence of stable \emph{exponential growth} of
original commits and files published over time.

Second, we study the number of \emph{different contexts} in which original code
artifacts re-appear over and over again, e.g., the same unmodified source code
file found in different commits, or the same commit distributed by different
repositories. By doing so we quantify the \emph{multiplication} of public
source code artifacts, addressing our second research question:
\begin{description}

\item[\RQii] to what extent the same source code artifacts, and in particular
  file and commits, can be found in different contexts (commits and
  repositories, respectively) in public source code?

\end{description}
We find evidence of a combinatorial explosion in the number of contexts in
which original source code artifacts appear, which is particular significant in
the multiplication of identical source code files across different commits.

In the last part of the paper, we explore the implications of such
multiplication on the problem of \emph{software provenance
  tracking}~\cite{Provenance2011, Godfrey15-provenance} for public source
code. We ask ourselves: is it feasible to keep track of all the different
contexts in which a given file or commit occur across the entire corpus?

To address this practical question we evaluate three different data models for
storing provenance information, which offer different space/time trade-offs.
We evaluate them on more than 40 years of public source code development
history and find that one of them---which we call the \emph{compact
  model}---allows to concisely track provenance across the entire body of
public source code, both today and in the foreseeable future.

\paragraph*{Paper structure}
we review related work in Section~\ref{sec:related} and address \RQi in
Section~\ref{sec:growth}; in Section~\ref{sec:provenancestudy} we attack \RQii,
studying the multiplication factor of original source code artifacts across
different contexts; provenance tracking representations are studied in
Section~\ref{sec:provenance}, leading to the compact model, which is
experimentally validated in Section~\ref{sec:validation}; threats to validity
are discussed in Section~\ref{sec:threats} before concluding in
Section~\ref{sec:conclusion}.

\paragraph*{Reproducibility note} given the sheer size of the \SWH archive
($\approx$200~TB and a $\approx$100~B edges graph), the most practical way to
reproduce the findings of this paper is to first obtain a copy of the official
\SWH Graph Dataset~\cite{msr-2019-swh-dataset} and then focus on the source
code revisions that we have analyzed for this paper. The full list of their
identifiers is available on Zenodo (DOI
\href{https://dx.doi.org/10.5281/zenodo.2543373}{10.5281/zenodo.2543373})
(20~GB); the selection criteria are described in Section~\ref{sec:growth}.

\section{Related Work}
\label{sec:related}

\begin{figure*}[t]
    \centering  \includegraphics[width=\linewidth]{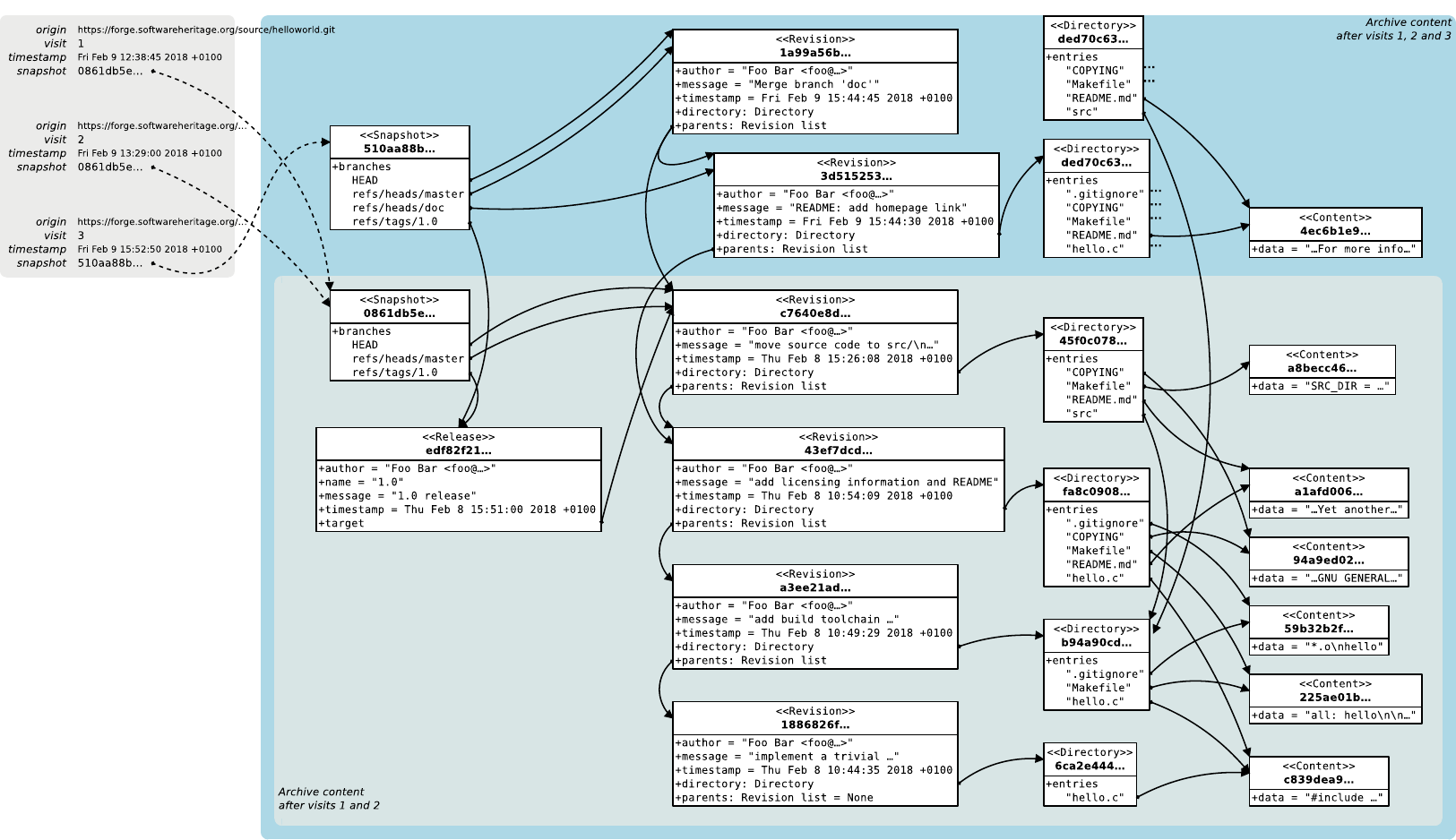}
  \caption{\SWH Merkle DAG with crawling information.}
  \label{fig:data-model-detailed}
\end{figure*}

The study of software evolution has been at the heart of software engineering
since the seminal ``Mythical Man Month''~\cite{BrooksMMM} and Lehman's
laws~\cite{LehmanLaw80}. The tidal wave of \FOSS, making available a growing
corpus of publicly available software, has spawned an impressive literature of
evolution studies. Some 10 years ago a comprehensive
survey~\cite{SurveyCrowston2008} showed predominance of studies on the
evolution of individual projects. Since then large scale studies have become
frequent and the question of how Lehman's laws need to be adapted to account
for modern software development has attracted renewed attention, as shown in a
recent survey~\cite{LawsEvolutionHerraizRRG13} that advocates for more
empirical studies to corroborate findings in the literature.

While Mining Software Research (MSR) research~\cite{hassan2008road} is
thriving, realizing large-scale empirical studies on software growth remains a
challenging undertaking depending on complex tasks such as collecting massive
amounts of source code~\cite{MockusAmassing09} and building suitable platforms
for analyzing them~\cite{dyer2013boa, Candoia2016}.
Hence, up to now, most studies have resorted to selecting relatively small
subsets\footnote{Some studies have analyzed up to a few million projects, but
  this is still a tiny fraction of all public source code.} of the full corpus,
using different criteria, and introducing biases that are difficult to
estimate.
For instance, an analysis of the growth of the Debian distribution spanning two
decades has been performed in~\cite{debsources-ese-2016}, observing initial
superlinear growth of both the number of packages and their size. But Debian is
a collection maintained by humans, so the number of packages in it depends on
the effort that the Debian community can consent.

A recent empirical study~\cite{HattonSG17} has calculated the compound annual
growth rate of over \num{4000} software projects, including popular \FOSS
products
as well as closed source ones. This rate is sensibly in the range of
1.20--1.22, corresponding to a \emph{doubling in size every 42 months}. In this
study, though, the size of software projects was measured using lines of code,
without discriminating between original contents and refactored or exogenous
code reused as-is from other projects.

Not many of these studies take into account the amount of code duplication
induced naturally by the now popular pull-request development
technique~\cite{gousios2014exploratory} and more generally by the ease with
which one can create copies of software components, even without forking them
explicitly.
The amount of exogenous code in a project can be extremely important, as shown
in~\cite{DejaVuVitek2017}, which analyzed over 4 million non-fork projects from
GitHub,
and showed that almost 70\% of the code consists of file-level exact
clones. This paints a very interesting picture of cloning in a subset of GitHub
at the time it was performed; it would be interesting to know how cloning
evolves over time, and how it impacts the growth of the global source code
corpus.

Software provenance tracking is an essential building block of several studies,
in particular on vulnerability tracking, license
analysis~\cite{GermanPGA09-siblings}, and reuse~\cite{ishio_software_2016}.
Provenance can be looked at different
granularities~\cite{BertillonnageGerman13}. On one end of the spectrum,
tracking the origin of code \emph{snippets} is useful when studying coding
patterns across repositories~\cite{AllamanisS13a, GermanPGA09-siblings}. On the
opposite end, tracking the origin of whole \emph{repositories} is useful when
looking at the evolution of forks or project popularity~\cite{Borges16}. In
between, tracking \emph{file}-level provenance has been for more than a decade
a key element of industrial tools for license compliance offered by companies
like BlackDuck, Palamida, Antelink, nexB, TripleCheck, or FossID, leading to
patent portfolios~\cite{rousseau_computer_2010}.

With few exceptions~\cite{Provenance2011}, though, file-level provenance has
received little attention in the research community.  We believe this is due to
the lack of a reference archive of public source code on which file-level
provenance tracking can be implemented once and then reused by other
researchers. In the final part of this paper we discuss the implications of our
findings about public source code and explore the feasibility of such
``provenance service'' approach, relying on \SWH as a proxy of public source
code.

\section{Public Source Code Growth}
\label{sec:growth}

Current development practices rely heavily on duplicating and reusing
code~\cite{gousios2014exploratory, DejaVuVitek2017}, which makes it non trivial
to estimate how much \emph{original} software is being produced: summing up the
usual metrics---such as number of source code files or revisions (also known as
\emph{commits})---across a wealth of software projects will inevitably end up
in counting the same original source code artifacts multiple times.

In this section we report on the first large scale analysis of the growth of
original software artifacts, in terms of revisions and contents, that we have
performed leveraging the fully-deduplicated data model that underlies \SWH,
briefly recalled below.

{\small \textbf{Terminological note}: we adopt in the following a
  technology-neutral terminology to refer to source code artifacts: we use
  ``content'' for ``[source code] file'' and ``revision'' for commit. The next
  subsection can be referred to for the intended meaning of those terms.}

\begin{figure*}[t!]
  \centering
    \includegraphics[width=\linewidth]{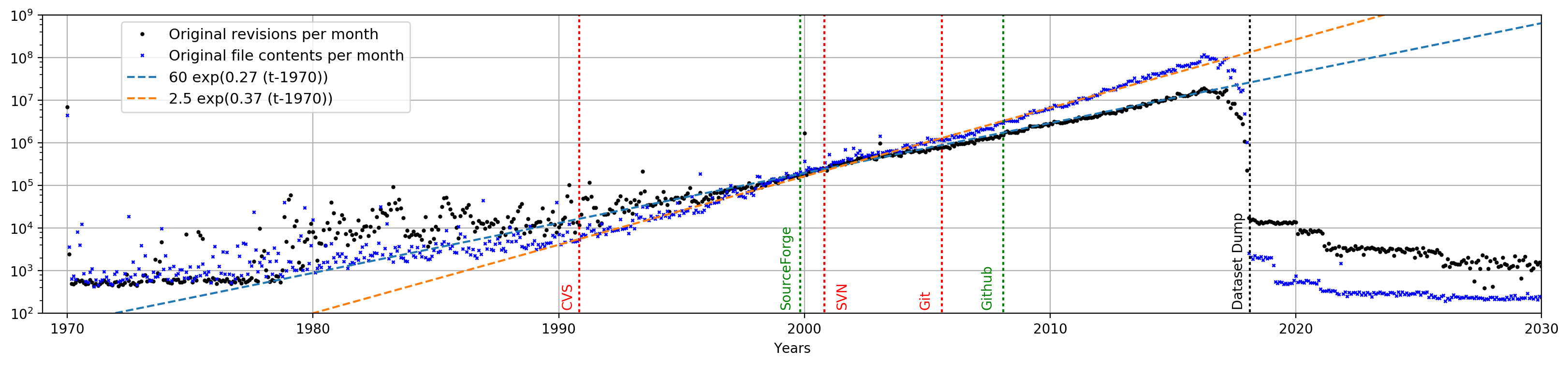}
    \caption{Global production of original software artifacts over time, in terms
    of never-seen-before revisions and file contents (lin-log scale).  Major
    events in the history of version control systems and development forges are
    materialised by vertical bars.}
  \label{fig:original-production}
\end{figure*}

\subsection{The \SWH data model}\label{sec:swh}

A toy yet detailed example of the \SWH data model is shown in
\figurename~\ref{fig:data-model-detailed}; full details can be found
in~\cite{swh-ipres-2017,swh-ipres-2018-doi}. The key principle is to deal with
code artifacts duplication by storing them in a single, huge Merkle direct
acyclic graph (DAG)~\cite{Merkle}, where every node is thoroughly
deduplicated. Different types of nodes are present in the graph:

\paragraph{Contents} raw file contents as byte sequences.  Contents are
anonymous; ``file names'' are given to them by directories and are context
dependent.

\paragraph{Directories} lists of named directory entries. Each entry can point
to content objects (``file entries''), to other directories (``directory
entries''), or even to other revisions (``revision entries''), capturing links
to external components like those supported by Git submodules and Subversion
externals). Each entry is associated to a name (i.e., a relative path) as well
as permission metadata and timestamps.

\paragraph{Revisions} (or \emph{commits}) point-in-time states in the
development history of a software project. Each revision points to the root
directory of the software source code at commit time, and includes additional
metadata such as timestamp, author, and a human-readable description of the
change.

\paragraph{Releases} (or \emph{tags}) particular revisions marked as noteworthy
by developers and associated to specific, usually mnemonic, names (e.g.,
version numbers or release codenames). Releases point to revisions and might
include additional descriptive metadata.

\paragraph{Snapshots} lists of pairs mapping development branch names (e.g.,
``master'', ``bug1234'', ``feature/foo'') to revisions or releases.
Intuitively each snapshot captures the full state of a development repository,
allowing to recursively reconstruct it if the original repository gets lost or
tampered with.

\emph{Deduplication} happens at node granularity for all source code artifacts:
each file content is stored exactly once and referred to via cryptographic
checksum key from multiple directories; each commit is stored once, no matter
how many repositories include it; up to each snapshot, which is stored once no
matter how many identical copies of repositories in exactly the same state
(e.g., pristine forks on GitHub) exist.

This arrangement allows to store in a uniform data model both specific versions
of archived software (pointed by release nodes), their full development
histories (following the chain of revision nodes), and development states at
specific points in time (pointed by snapshot nodes).

In addition to the Merkle DAG, \SWH stores \emph{crawling information}, as
depicted in the top left of \figurename~\ref{fig:data-model-detailed}. Each
time a source code origin is visited, its full state is captured by a snapshot
node (possibly reusing a previous snapshot node, if an identical repository
state has been observed in the past) plus a 3-way mapping between the origin
(as an URL), the visit timestamp, and the snapshot object, which is then added
to an append-only journal of crawling activities.

\subsection{Key figures on the \SWH dataset}

At the time we used it for this paper, the \SWH archive was the largest
available corpus of public source code~\cite{swh-ipres-2017, swh-cacm-2018},
encompassing:
\begin{itemize}
\item a full mirror of GitHub, constantly updated
\item a full mirror of Debian packages, constantly updated
\item a full import of the Git and Subversion repositories hosted on Google
  Code at shutdown time
\item a full import of Gitorious at shutdown time
\item a one-shot import of all GNU packages (\emph{circa} 2016)
\end{itemize}

\begin{table}
  \caption{Graph characteristics of the reference dataset: a \SWH archive copy
    as of February 13th, 2018.}
  \label{tab:ref-dataset}
  \centering
  \subfigure[archive coverage]{
    46.4 M software origins
  }\\
  \subfigure[nodes]{
    \begin{tabular}{l|l}
      \multicolumn{1}{c|}{\bf node type}
      & \multicolumn{1}{c}{\bf quantity}
      \\\hline
      content   & 3.98 B \\
      revision  & 943 M \\
      release   & 6.98 M \\
      directory & 3.63 B \\
      snapshot  & 49.9 M \\
      \hline
      \it total & 8.61 B \\
    \end{tabular}
  }
  \subfigure[edges]{
    \begin{tabular}{l|l}
      \multicolumn{1}{c|}{\bf edge type}
      & \multicolumn{1}{c}{\bf quantity}
      \\\hline
      revision $\to$ directory  & 943 M \\
      release $\to$ revision    & 6.98 M \\
      snapshot $\to$ release    & 200 M \\
      snapshot $\to$ revision   & 635 M \\
      snapshot $\to$ directory  & 4.54 K \\
      directory $\to$ directory & 37.3 B \\
      directory $\to$ revision  & 259 M \\
      directory $\to$ file      & 64.1 B \\
      \hline
      \it total                 & 103 B   \\
    \end{tabular}
  }
\end{table}

For this paper we used the state (called \emph{reference dataset} in the
following) of the full \SWH archive as it was on February 13th, 2018. In terms
of raw storage size, the dataset amounts to about 200~TB, dominated by the size
of content objects. As a graph, the DAG consists of $\approx$9~B nodes and
$\approx$100~B edges, distributed as shown in Table~\ref{tab:ref-dataset}; note
how this corpus is orders of magnitudes larger than previously analyzed
ones~\cite{debsources-ese-2016, DejaVuVitek2017, MLgitarchive18}.

\subsection{Evolution of original revisions and file contents}

We have analyzed the entire reference dataset (see
Table~\ref{tab:ref-dataset}), processing revisions in increasing timestamps
order, and keeping track for each file content the timestamp of the
\emph{earliest} revision that contains it, according to the commit timestamp.
A revision is \emph{original} if the combination of its properties (or,
equivalently, its identifier in the Merkle DAG) has never been encountered
before.
Results are shown in \figurename~\ref{fig:original-production}. They provide
very rich information, answering \RQi for both revisions and file contents.

We discuss first a few outliers that jump out. Data points at the \emph{Unix
  epoch} (1/1/1970) account for 0.75\% of the dataset and are clearly
over-represented. They are likely due to forged revision timestamps introduced
when converting across version control systems (VCSs). This is probably also
the main reason behind revisions with timestamps in the ``future'', i.e., after
the dataset timestamp (0.1\% of the dataset). The sharp drop before the dataset
timestamp is a consequence of the lag of \SWH crawlers w.r.t.~its data sources.

Focusing on the core part of the figure we remark that in the early years,
before the introduction of forges and advanced VCSs, the number of revisions is
relatively small (tens to hundreds of thousands only), and their evolution is
rather irregular.

After the creation of the first popular forge, SourceForge (1999), we observe
on the other hand a remarkably regular exponential growth lasting twenty years.
For original revisions, growth can be accurately approximated by the fit line
$60 e^{0.27(t-1970)}$; at this rate \textbf{the amount of original revisions in
  public source code doubles every $\approx$30 months}. For original contents,
growth is accurately approximated by the fit line $2.5 e^{0.37(t-1970)}$; at
this rate \textbf{the amount of original public source code files doubles every
  $\approx$22 months}.

This information is precious to estimate the resources needed for
\emph{archiving} publicly developed software: taking into account the long term
evolution of storage costs\footnote{see, e.g.,
  \url{https://hblok.net/blog/storage/}} this growth looks managable, provided
that deduplication is applied.  The sustainability of provenance tracking
remains potentially challenging, though, because artifact \emph{occurrences} in
different contexts cannot be deduplicated. We will quantify source code
artifact multiplication in the next section.

The growth rate of original contents and revisions suggests that both the
production of source code and its derived graphs are interesting evolving
complex networks~\cite{albert2002statistical, dorogovtsev2002evolution}. Their
nature---scale-free or not---as well as the role of phenomena like preferential
attachment in the growth dynamics between edges and nodes, potentially leading
to accelerating growth~\cite{albert2002statistical}, are important subjects for
further investigation.

Finally, we remark that the \emph{difference} in the growth rates of original
revisions and original file contents means that over the past twenty years
\textbf{the average number of original file contents per revision has been
  doubling every $\approx$7 years}. Whether this comes from the availability of
better tools that can easily handle large commits, from different development
practices, or other causes is another interesting open question.

\section{Public Source Code Multiplication}
\label{sec:provenancestudy}

\begin{figure}
  \centering
  \includegraphics[width=\figstretch\linewidth]{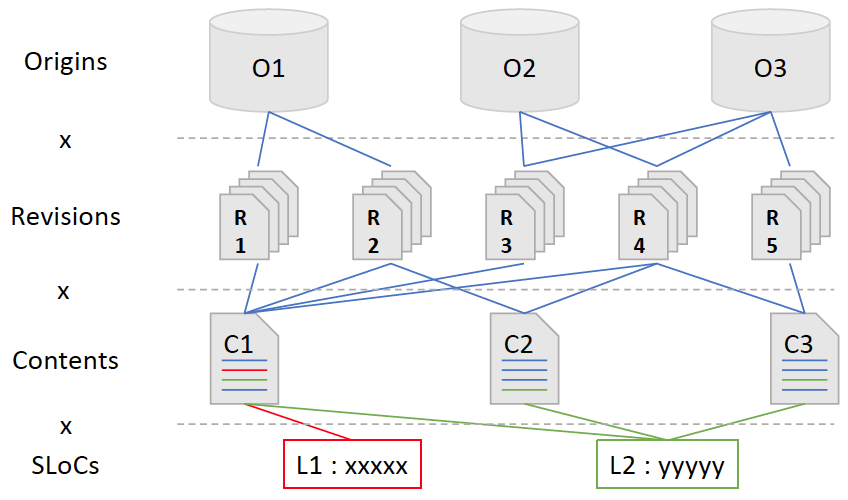}
  \caption{The three layers of multiplication in public source code: SLOCs
    occurring in source code files (contents), contents occurring in commits
    (revisions), revisions found at different distribution places (origins).}
  \label{fig:multiplication}
\end{figure}

We now look into \emph{public source code multiplication}, i.e., how often the
same artifacts (re-)occur in different contexts.
\figurename~\ref{fig:multiplication} depicts the three layers of this
phenomenon: a given line of code (SLOC) may be found in different source code
files; a given file content may appear in different revisions (e.g., different
commits in the same repository); and a given revision may be found at multiple
origins (e.g., the same commit distributed by multiple repositories and source
packages).

To study this phenomenon and answer \RQii we perform in the following focused
analyses on the \SWH Merkle DAG. They will lead to quantitatively evaluate the
\emph{multiplication factor} of source code artifacts at each multiplication
layer of \figurename~\ref{fig:multiplication}.

\subsection{Content multiplication factor}
\label{sec:contentdup}

\begin{figure}
  \centering
  \includegraphics[width=\figstretch\linewidth]{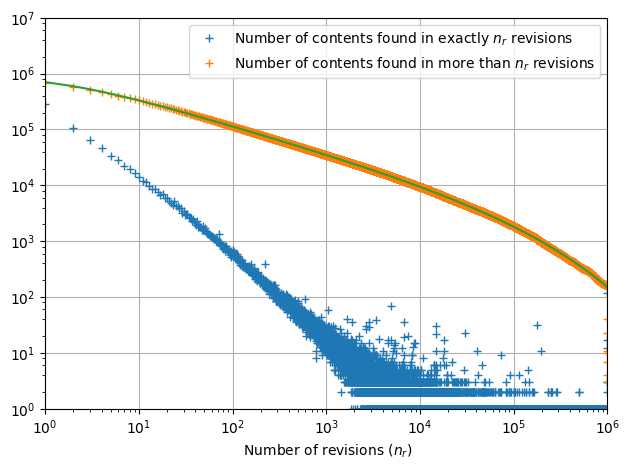}
  \includegraphics[width=\figstretch\linewidth]{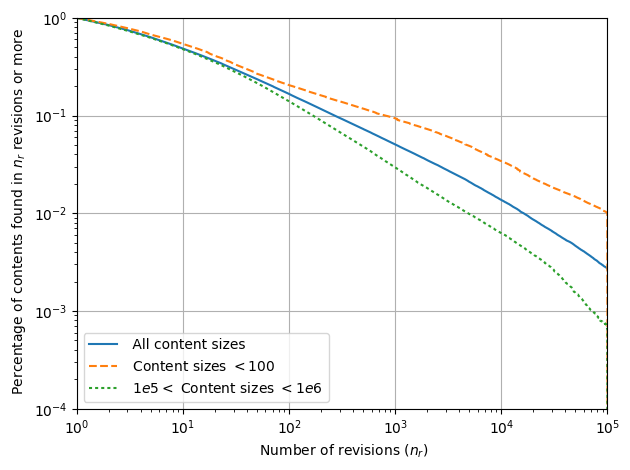}
  \caption{Top: cumulative (upper curve) and simple (lower curve)
    multiplication factor of unique file contents across unique revisions.
    Bottom: normalized cumulative content multiplication factor for the same
    sample (solid line) and two random samples of about 1~M contents each, with
    content sizes up to 100~bytes (dashed line) and between $10^5$ and $10^6$
    bytes (dotted line).}
  \label{fig:content-duplication}
\end{figure}

In order to assess the \emph{content} multiplication factor, i.e., the amount
of duplication of file contents among revisions, we took a random sample of
about 1 million unique contents (all contents whose hash identifiers start with
\texttt{aaa}. For each content in that sample we counted how many revisions
contain it in the reference dataset. The resulting distribution of the
multiplication factor is shown in the upper part of
\figurename~\ref{fig:content-duplication}, together with the corresponding
cumulative distribution.

Looking at the cumulative distribution it jumps out that the average
multiplication factor is very high. It exhibits a characteristic decreasing
power law ($\alpha\simeq -1.5$), only limited by an exponential cut-off.  There
are hence over a hundred of thousand contents that are duplicated more than one
hundred times; tens of thousand contents duplicated more than a thousand times;
and there are still thousands of contents duplicated \emph{more than a hundred
  thousands times}! Space-wise, keeping track of all the occurrences of the
content$\to$revision layer of \figurename~\ref{fig:multiplication} is a highly
nontrivial task.

We did not resist investigating the side question of whether the \emph{size} of
a file content impacts the multiplication factor. We hence took two new random
samples of about 1 million contents each, one with content sizes up to 100
bytes and one with sizes between $10^5$ and $10^6$ bytes, and performed the
same analysis as for the previous sample.

The resulting normalized cumulative multiplication factors are shown on the
bottom of \figurename~\ref{fig:content-duplication}. We can see that \emph{the
  multiplication factor of small contents is much higher} than that of
average-sized and large contents.  Hence, keeping track of the
content$\to$revision occurrences only for files larger than, say, 100 bytes, is
a significantly simpler problem than its fully general variant. Omitting small
files is indeed a technique often used by state-of-the-art industry solutions
for software provenance tracking: we provide evidence on why it is effective
(at the expense of completeness).

\subsection{SLOC length and multiplication factor}

\begin{figure}
  \centering
  \includegraphics[width=\figstretch\linewidth]{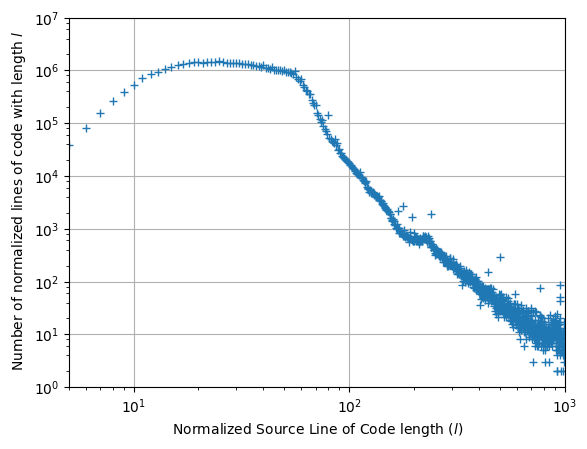}
    \caption{Distribution of normalized SLOC lengths in a sample of 2.5~M
    contents that appear at least once with \texttt{.c} extension.}
  \label{fig:loc-size}
\end{figure}

\begin{figure}
  \centering
  \includegraphics[width=\figstretch\linewidth]{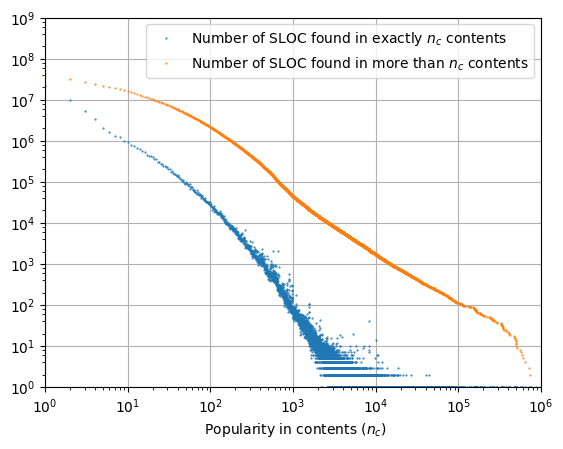}
  \caption{Multiplication factor of normalized SLOCs as the number of unique
    contents they appear in. Dataset: same of \figurename~\ref{fig:loc-size}.
  }
  \label{fig:loc-duplication}
\end{figure}

We now turn our attention to the bottom layer of
\figurename~\ref{fig:multiplication}: SLOC$\to$content. Since lines of code are
hardly comparable across languages, we focused on the C language, which is
well-represented in the corpus. We took a random sample of $\approx$11.4~M
unique contents occurring in revisions between 1980 and 2001, and selected from
it contents that appear at least once with \texttt{.c} extension and with sizes
between $10^2$ and $10^6$ bytes, obtaining $\approx$2.5~M contents. We then
split contents by line and, to remove equivalent formulations of the same SLOC,
\emph{normalized} lines by removing blanks and trailing \texttt{";"}
(semicolon). We obtained $\approx$64~M normalized SLOCs.

The multiplication factor of SLOCs across unique contents is shown in
\figurename~\ref{fig:loc-duplication}. We observe a much faster decrease
w.r.t.~the multiplication factor of contents in releases ($\alpha\simeq -2.2$),
hinting that keeping track of SLOCs$\to$content occurrences may be less
problematic than content$\to$revision.

We also computed the distribution of normalized SLOC lengths between 4 and
1000, shown in \figurename~\ref{fig:loc-size}. We observe that lines with
length 15 to 60 normalized characters are the most represented, with a fairly
stable presence within that range, and a steep decrease for longer lines.
Hence, for SLOC$\to$content occurrences there does not seem to exist any
obvious length-based threshold that would reduce their amount.

\subsection{Origin size and multiplication factor}

\begin{figure}
  \centering
  \includegraphics[width=\figstretch\linewidth]{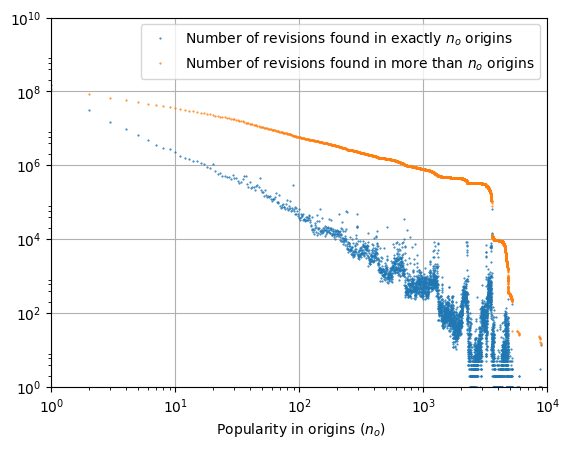}
  \caption{Duplication of revisions across origins.}
              \label{fig:origin-duplication}
\end{figure}

Finally we look into the revision$\to$origin layer of
\figurename~\ref{fig:multiplication}. To that end we took a sample of
$\approx$12\% of the origins, which contain $\approx$29\% of the revisions (or
about 5.4~M origins and 272.5~M revisions) and replicated the previous study of
content duplication onto revisions. Results are shown in
\figurename~\ref{fig:origin-duplication}.

Revision multiplication shows an erratic behavior near the end of the range,
but decreases steadily before that, and way more steeply ($\alpha\simeq -1.9$)
than it was the case for content$\to$revision multiplication (see
\figurename~\ref{fig:content-duplication} for comparison): the multiplication
factor of revisions in origins is way smaller than that of contents in
revisions.

While this result is sufficient to assess the respective impact on public
source code multiplication of the considered layers, we dug further into origin
sizes to better understand \emph{which} origins participate into
revision$\to$origin multiplication.

We have considered two different measures of origin size. One that simply
counts the number of revisions found at each origin. Another that associates
revisions found at multiple origins \emph{only to the origin that contains the
  largest number of revisions}, and the count them as before. When a project is
forked, the second measure would always report a revision as belonging to the
fork with the most active development, which is a good approximation of the
``most fit fork'', while stale forks would decay. This measure has many good
properties: it will follow forks that resurrect projects abandoned at their
original development places, it does not rely on platform metadata for
recognizing forks, and is hence able to recognize \emph{exogenous forks} across
unrelated development platforms (e.g., GitHub-hosted forks of the Linux kernel
that is not natively developed on GitHub).

\begin{figure}
  \centering
  \includegraphics[width=\figstretch\linewidth]{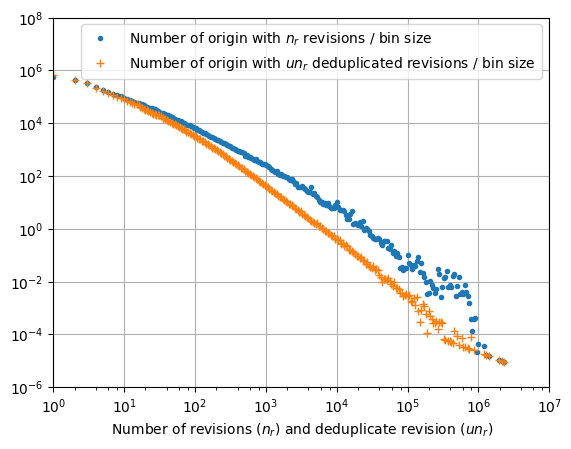}
    \caption{Distribution of origin size as the number of revisions they host.}
                \label{fig:origin-size}
\end{figure}

\figurename~\ref{fig:origin-size} shows the impact that the ``most fit fork''
measure has on the number of revision$\to$origin occurrences. Starting with
relatively small repositories, ($\approx$100 revisions) the number of
occurrences to track is lower than for the simpler measure, with a difference
growing up to a full order of magnitude for repositories hosting 10~K
revisions.

\section{Compact Provenance Modeling}
\label{sec:provenance}

We now consider the problem of tracking software provenance across a corpus as
large (and as fast growing) as all public source code. In short, the goal is to
keep track of all the different places (contents, revisions, origins) in which
any given source code artifact (SLOC, content, revision) can be found---more
detailed requirements are given below in Section~\ref{sec:requirements}.

What are the implications of our findings on public source code growth and
multiplication, on the \emph{feasibility} of maintaining such a complete
provenance index?  An important fact that emerges from the analyses is that,
size-wise, the most challenging part is the layer associating file contents to
all the revisions they appear in, because contents are duplicated across
revisions much more than revisions across origins.

Hence in the following we will focus on concisely representing the
content$\to$revision mappings; the revision$\to$origin ones will be a
straightforward and fully modular addition.  SLOC-level provenance tracking is
left as future work.

\subsection{Requirements}
\label{sec:requirements}

\paragraph{Supported queries} At least two queries should be supported: first
occurrence and all occurrences. The \emph{first occurrence} query shall return
the earliest occurrence of a given source code artifact in any context,
according to the revision timestamp. The \emph{all occurrences} query will
return all occurrences. The two queries answer different use cases: first
occurrence is useful for prior art assessment and similar intellectual property
needs; all occurrences is useful for impact/popularity analysis and might be
used to verify first occurrence results in case of dubious timestamps.

\paragraph{Granularity} It should be possible to track the provenance of source
code artifacts at different granularities including at least file contents and
revisions.

\paragraph{Scalability} It should be possible to track provenance at the scale
of at least \SWH and keep up with the growth rate of public source code. Given
that the initial process of populating provenance mappings might be onerous,
and that some use cases require fresh data (e.g., impact/popularity), we also
require \emph{incrementality} as part of scalability: the provenance index must
support efficient updates of provenance mappings as soon as source code
artifacts (old or new) are observed in new contexts.

\paragraph{Compactness}
It should be possible to store and query provenance information using
state-of-the-art consumer hardware, without requiring dedicated hardware or
expensive cloud resources.

\paragraph{Streaming} For the \emph{all occurrences} query a significant
performance bottleneck is the transfer time required to return the potentially
very large result. A viable provenance solution should hence allow to return
results incrementally, piping up the rest for later.

\subsection{Provenance data models}

\begin{figure}
  \centering
  \subfigure[flat model]{
    \includegraphics[width=0.6\linewidth]{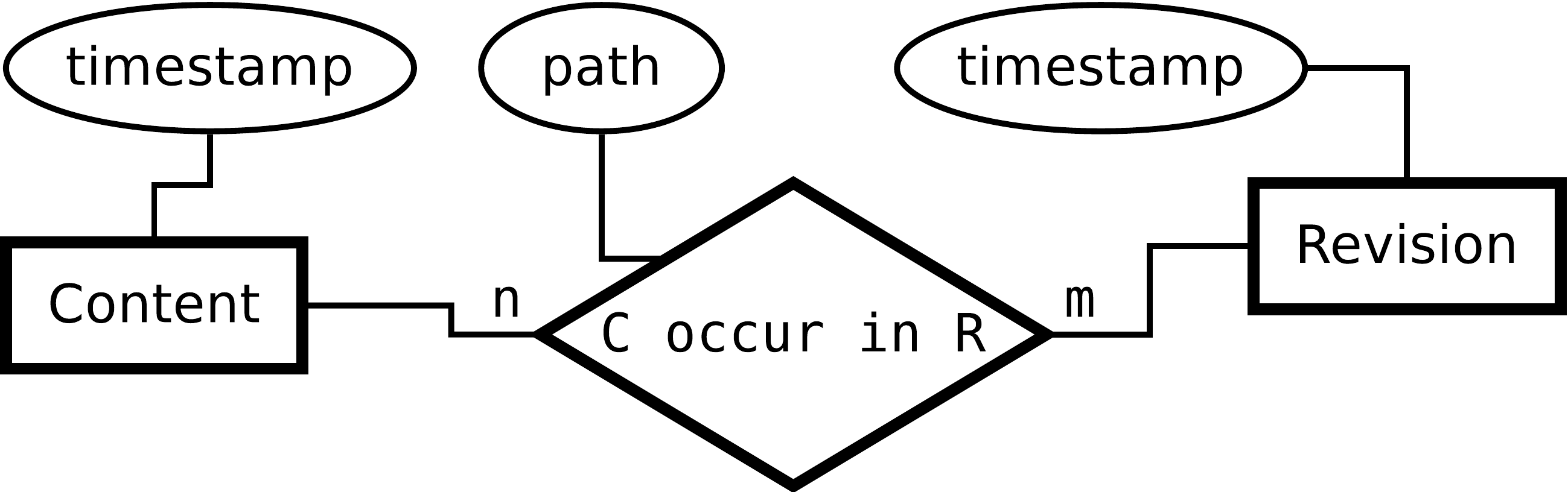}
    \label{fig:flat-model}
  }
  \subfigure[recursive model]{
    \includegraphics[width=\linewidth]{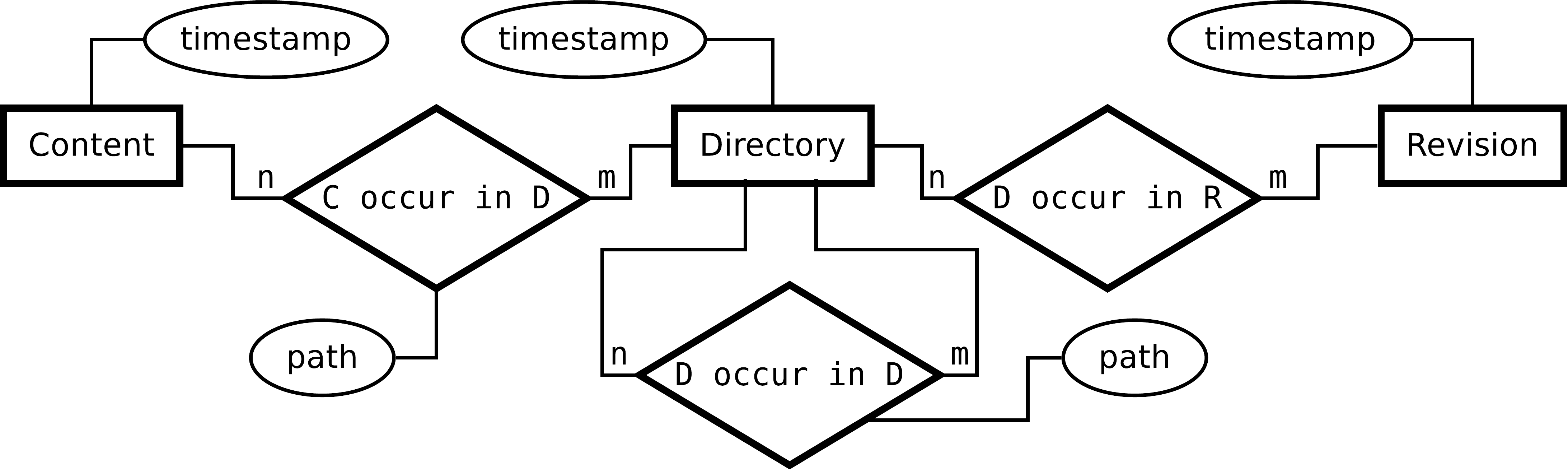}
    \label{fig:recursive-model}
  }
  \subfigure[compact model]{
    \includegraphics[width=\linewidth]{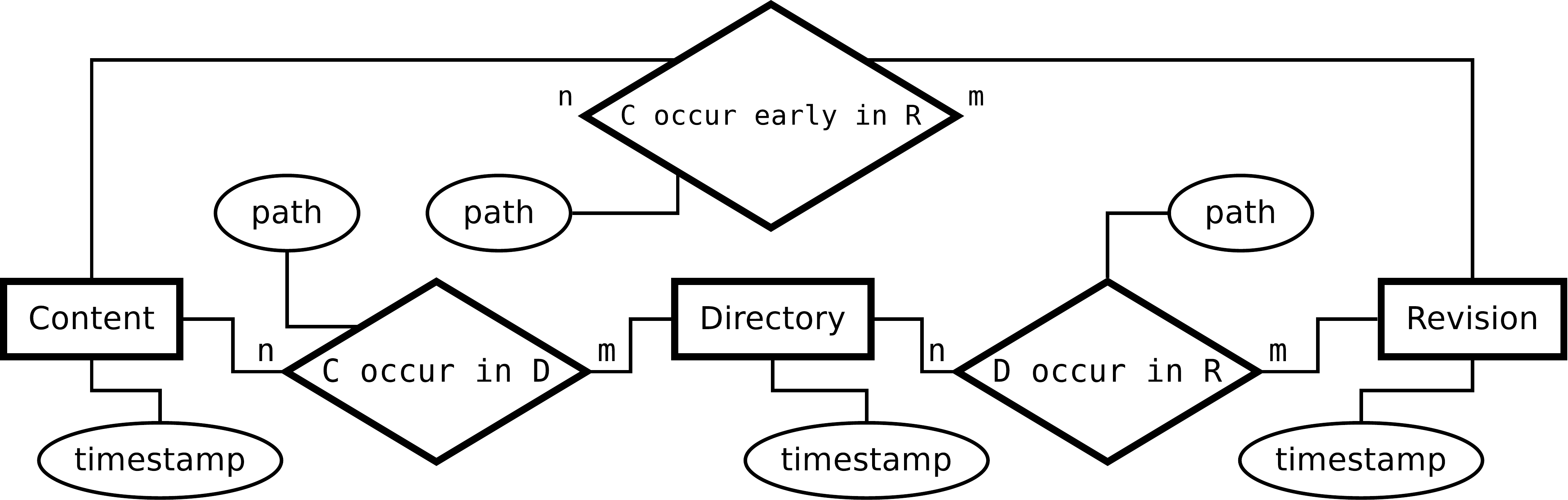}
    \label{fig:compact-model}
  }
  \caption{Provenance tracking models, entity-relationship (E-R) views}
  \label{fig:provenance-models}
\end{figure}

We study three different data models for provenance tracking, that we call
respectively \emph{flat}, \emph{recursive}, and \emph{compact}. Their
Entity-Relationship (E-R) representations are shown in
\figurename~\ref{fig:provenance-models}.

\paragraph{Flat model}
this is our baseline for tracking provenance, shown in
\figurename~\ref{fig:flat-model}. In this model provenance mappings are
``flattened'' using a single \errel{C(ontent) occur in R(evision)} relation,
that also keeps track of file paths relatively to the root directory of the
associated revision. The cardinality of \errel{C occur in R} is n-m (rather
than 1-n), because the same content might appear multiple times in a given
revision at different paths. Each revision carries as attribute the revision
timestamp, in order to answer the question of \emph{when} the occurrence
happened. Each content carries as attribute the timestamp of its earliest
occurrence, i.e., the minimum timestamps among all associated revisions.

Given suitable indexing on content identifiers (e.g., using a B-tree), the flat
model adds no read overhead for the all occurrences query. Same goes for first
occurrence, given suitable indexing on timestamp attributes, which is required
to retrieve path and revision.

Updating provenance mappings when a new revision comes in requires traversing
the associated directory in full, no matter how many sub-directories or
contents in it have been encountered before, and adding a relationship entry
for each of its nodes.

\paragraph{Recursive model}
while the flat model shines in access time at the expenses of update time and
compactness, the recursive model shown in \figurename~\ref{fig:recursive-model}
does the opposite. It is intuitively a ``reverse'' Merkle DAG representation,
which maps contents to directories and directories to revisions.

Each entity has a timestamp attribute equal to the timestamp of the earliest
revision in which the entity has been observed thus far. When processing an
incoming revision $r_{t_2}$ (with timestamp $t_2$) it is no longer necessary to
traverse in full the associated directory: if a node $n$ is encountered that
is already present in the model with a timestamp $t_1<t_2$, recursion can stop because the
subtree rooted at $n$, which is already present due to the Merkle DAG properties,
has already been labeled with timestamps earlier than $t_2$ and needs not be updated;
we just need to add an entry in the corresponding occurrence table for $n$ with timestamp $t_2$.

Thanks to the sharing offered by the directory level, the recursive model is as
compact as the original Merkle structure, with no flattening involved. The all
occurrences query is slow in this model though, as for each content we need to
walk up directory paths before finding the corresponding revisions. Response
time will hence depend on the average directory depth at which queried contents
will be found. First occurrence is faster, but still incurs some read overhead:
given a content we have to walk up all directories and then lookup the
corresponding revisions whose timestamps equate the timestamp of the content
being queried.

\paragraph{Compact model}
\figurename~\ref{fig:compact-model} shows a compromise version between the flat
and recursive models, which is both storage-compact and capable of quickly
answering the required queries. The tables for the content, directory, and revision entities
are progressively populated as the structure is built, with a timestamp attribute denoting the earliest
known occurrence, as before. To understand how the compact model is built and used we introduce the following
notion:
\begin{definition}[Isochrone subgraph] \it   given a partial provenance mapping $\mathcal{P}$ associating a timestamp of
  first occurrence to each node in a Merkle DAG, the \emph{isochrone subgraph}
  of a revision node $R$ (with timestamp $t_R$) is a subgraph rooted at $R$'s
  directory that only contains directory nodes whose timestamps in
  $\mathcal{P}$ are equal to $t_R$.
\end{definition}

Intuitively, when processing revisions chronologically to update the entity tables and the provenance
mappings, the isochrone subgraph of a revision starts with its root directory
and extends through all directory nodes containing never-seen-before source
code artifacts. Due to Merkle properties each directory containing at least one
novel element is itself novel. Everything outside the isochrone subgraph has
been observed before, in at least one previously processed revision.

Given this notion, the upper part of the compact model (\errel{C occur early in
  R} in \figurename~\ref{fig:compact-model}) is filled with one entry for each
content attached to any directory in the isochrone subgraph. As a consequence
of this, the first occurrence of any given content will always be found in
\errel{C occur early in R} although other occurrences---depending on the order
in which revisions are processed to update provenance mappings---may also be
found there.

The relation \errel{D occur in R} is filled with one entry, pointing to the
revision being processed, for each directory \emph{outside} the isochrone
subgraph that is referenced by directories \emph{inside} it, i.e., \errel{D
  occur in R} contains one entry for each directory$\to$directory edge crossing
the isochrone frontier. Finally, the relation \errel{C occur in D} is filled
with one entry for each content (recursively) referenced by any directory added
to the \errel{D occur in R} relation.

Filling the compact model is faster than the flat model: when we reach a
directory $d$ at the frontier of an isochrone subgraph, we only need to
visit it in full the first time, to fill \errel{C occur in D}, and we
need not visit $d$ again when we see it at the frontier of another
isochrone subgraph in the future.

It is slower than the recursive model case, though, as we still need
to traverse the isochrone subgraph of each revision. Read overhead for first
occurrence is similar to the flat model: provided suitable indexing on
timestamps we can quickly find first occurrences in \errel{C occur early in R}.
Read overhead for all occurrences is lower than the recursive model because all
content occurrences will be found via \errel{C occur in D} without needing to
recursively walk up directory trees, and from there directly linked to
revisions via \errel{D occur in R}.

\subsection{Discussion}

Intuitively, the reason why the compact model is a good compromise is that we
have many revisions and a very high number of file contents that occur over and
over again in them, as discussed in Section~\ref{sec:contentdup}.  Consider now
two extreme cases: (1) a set of revisions all pointing to the same root
directory but with metadata differences (e.g., timestamp or author) that make
all revisions unique; (2) a set of revisions all pointing to different root
directories that have no file contents or (sub)directories in common.

In case (1) the flat model would explode in size due to maximal duplication.
The recursive model will need just one entry in \errel{D occur in R} for each
revision. The compact model remains small as the earliest revision will be
flattened (via \errel{C occur early in R}) as in the flat model, while each
additional revision will add only one entry to \errel{D occur in R} (as in the
recursive model).

In case (2) the flat model is optimal in size for provenance tracking
purposes, as there is no sharing. The recursive model will have to store all
deconstructed paths in \errel{D occur in D}. The compact model will be
practically as small as the flat model: all revisions are entirely isochrones,
so the \errel{C occur early in R} relation will be the same as the \errel{C occur in R} of the flat model, and the only extra item is the \errel{Directory} table.

Reality will sit in between these two extreme cases, but as the compact model
behaves well in both, we expect it to perform well on the real corpus too. The
experimental evaluation reported in the next section validates this intuition.

\section{Evaluation}
\label{sec:validation}

To compare the size requirements of the provenance data models described in
Section~\ref{sec:provenance}, we have monitored the growth of each model while
processing incoming revisions to maintain provenance mappings up to date.

Specifically, we have processed in chronological order revisions from the
reference dataset with timestamps strictly greater than the Unix epoch (to
avoid the initial peak of forged revisions discussed in
Section~\ref{sec:growth}) and up to January 1st, 2005, for a total of
$\approx$38.2~M revisions. For each revision we have measured the number of
entities and relationship entries according to the model definitions, that is:

\paragraph{Flat model} one entity for each content and revision; plus one
\errel{\small C occur in R} entry for each content occurrence

\paragraph{Recursive model} as it is isomorphic to the Merkle DAG, we have
counted: one entity for each content, directory, and revision; plus one
relationship entry for each revision$\to$directory, directory$\to$directory,
and directory$\to$content edge

\paragraph{Compact model} after identifying the isochrone subgraph of each
revision, we counted: one entity for each content and revision, plus one entity
for each directory outside the isochrone graph referenced from within; as well
as one relationship entry for each content attached to directories in the
isochrone graph (\errel{C occur early in R}), one \errel{D occur in R} entry
for each directory$\to$directory edge crossing the isochrone frontier, and one
\errel{C occur in D} entry for each content present in directories appearing in
\errel{D occur in R}.

Processing has been done running a Python implementation of the above
measurements on a commodity workstation (Intel Xeon 2.10GHz, 16 cores, 32 GB
RAM), parallelizing the load on all cores. Merkle DAG information have been
read from a local copy of the reference dataset, which had been previously
mirrored from \SWH. In total, revision processing took about 4 months, largely
dominated by the time needed to identify isochrone subgraphs.

\begin{table}
  \centering
  \caption{Size comparison for provenance data models, in terms of entities and
    relationship entries. Same dataset of \figurename~\ref{fig:model-sizes}.}
  \label{tab:model-sizes}
    \begin{tabular}{l|r|r|r}
    & \multicolumn{1}{c|}{\textbf{Flat}}
    & \multicolumn{1}{c|}{\textbf{Recursive}}
    & \multicolumn{1}{c}{\textbf{Compact}} \\
    \hline
    entities     & \num{80118995}     & \num{148967553}   & \num{97190442}    \\
                 & rev: 38.2 M        & rev: 38.2 M       & rev: 38.2 M       \\
                 & cont: 41.9 M       & cont: 31.9 M      & cont: 31.9 M      \\
                 &                    & dir: 68.8 M       & dir: 17.1 M       \\
    \hline
    rel.~entries & \num{654390826907} & \num{2607846338}  & \num{19259600495} \\
                 &                    & cont--dir: 1.29 B & cont--dir: 13.8 B\\
                 &                    & dir--rev: 38.2 M  & dir--rev: 2.35 B  \\
                 &                    & dir--dir: 1.28 B  & cont--rev: 3.12 B  \\
    \hline\hline
    rel.~ratios
     & \multicolumn{1}{c|}{$\frac{flat}{compact}$ = 34.0}
     & \multicolumn{1}{c|}{$\frac{flat}{rec.}$ = 251}
     & \multicolumn{1}{c}{$\frac{compact}{rec.}$ = 7.39}
    \\
  \end{tabular}
\end{table}

\begin{figure}
  \centering
  \includegraphics[width=\figstretch\linewidth]{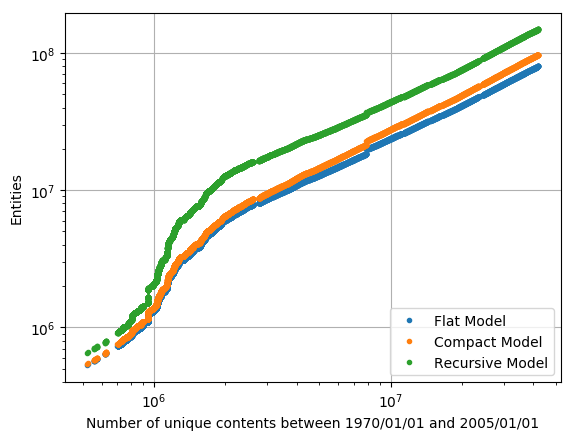}
  \includegraphics[width=\figstretch\linewidth]{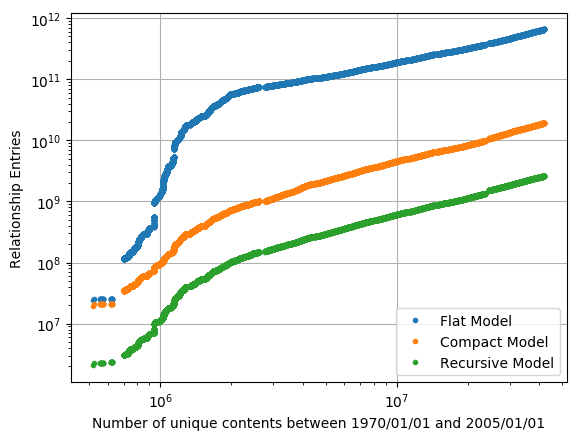}
  \caption{Evolution over time of the sizes of different provenance data
    models, in terms of entities (top) and relationship entries (bottom). Data
    for \SWH revisions up to 2005-01-01, excluding Unix epoch.
  }
  \label{fig:model-sizes}
\end{figure}

Final sizes, measured in terms of entities and relationship entries are given
in Table~\ref{tab:model-sizes}. They show, first, that the amount of
relationship entries dominate that of entities in all models, from a factor 18
(recursive model) up to a factor \num{8000} (flat). Dealing with mappings
between source code artefacts remains the main volumetric challenge in
provenance tracking. As further evidence of this, and as a measure of the
overall amplitude of provenance tracking for all public source code, we have
also computed the number of relationship entries for the flat data model
\emph{on the full reference dataset}, obtaining a whooping $8.5\cdot 10^{12}$
entries in \errel{C occur in R}.

Second, sizes show that the Merkle DAG representation, isomorphic to the
recursive model, is indeed the most compact representation of provenance
information, although not the most efficient one to query. The compact model is
the next best, 7.39 times larger than the recursive model in terms of
relationship entries. The flat model comes last, respectively 251 and 34 times
larger than recursive and compact.

\figurename~\ref{fig:model-sizes} shows the evolution of model sizes over time,
as a function of the number of unique contents processed thus far. After an
initial transition period, trends and ratios stabilize making the outlook of
long-term viability of storage resources for the compact model look good.

Furthermore, the comparison between the compact (orange line) and flat (blue)
model shows that, at the cost of a small increase in the number of entities,
the compact model performs much better in terms of relationship entities. And
even if, in terms of entities a small divergence can be observed over time
($1/10$ of an order of magnitude), the gain in terms of relationship entries
makes it worthwhile (1.5 orders of magnitude).

In order to relate these figures to real-world storage requirements, we have
also filled a MongoDB-based implementation of the compact model---including all
attributes of \figurename~\ref{fig:compact-model} and needed indexes---while
processing revisions to perform the above measurements. Extrapolating the final
MongoDB size to the full reference dataset we obtain an on-disk size of
13~TB. While large, such a database can be hosted on a consumer workstation
equipped with $\approx$\num{4000}\$ of SSD disks, without having to resort to
dedicated hardware or substantial investments in cloud resources.  Using the
compact model, universal source code provenance tracking can lay at the
fingertips of every researcher and industrial user!

\section{Threats to Validity}
\label{sec:threats}

\paragraph{Internal validity}
The main concern for internal validity is that we did not have the resources
available to perform all estimates and experiments on the full \SWH archive.
While our main growth results are measured on the full reference dataset, other
results are extrapolated from smaller subsets. To counter potential bias, we
have used random samplings and sizeable samples.

When comparing provenance data models we have quantitatively estimated sizes,
but only qualitatively estimated read overhead---rather than benchmarking it in
production---in order to remain technology-neutral.

Finally, we have trusted commit timestamps to determine first occurrences, even
if it commit timestamps can be forged. This approach is consistent with
previous software evolution studies that consider timestamp forging a marginal
phenomenon. We also remark that for evaluating the performance of the
provenance model we only need to single out \emph{a} ``first'' occurrence, no
matter how it is determined.

\paragraph{External validity}
While \SWH does not cover all existing free/open source software, it is the
largest source code archive in the world and spans the most popular code
hosting and development platforms. We therefore consider that this is the best
that can be done at present.

Finally, we acknowledge the habit of using \emph{software} development
platforms for collaboration tasks other than software development (e.g.,
collaborative writing), particularly on GitHub, but we did not try to filter
out non-software projects. On the one hand we expect software development to be
the dominant factor, and on the other hand non-software projects might still
contain interesting code snippets that are worth tracking. Also, as
demonstrated in the paper, it is not \emph{necessary} to filter out
non-software project in order to build a practical provenance tracking
solution.

\section{Conclusion}
\label{sec:conclusion}

The emergence of \SWH as a comprehensive archive of public source code,
spanning tens of millions of software projects over more than 40 years, enables
analysis of the evolution of software development at a novel scale.

The first contribution of this paper is a quantitative analysis of the growth
of public software development, factoring out exact code clones. Since the
advent of version control systems, the production of unique original revisions
doubles every 30 months, and the production of unique original file is even
faster, doubling every 22 months. Besides confirming the perceived overall
growth of the public software ecosystem, these results open up a wealth of new
research questions.

The second contribution is a quantitative assessment of the amount of
duplication of both original file contents across different commits and of
original commits across different software origins, gaining precious
preliminary insights into the deep structure of public software development and
distribution.

The third and final contribution is the development and comparison of three
data models designed to answer the software provenance questions of ``what are
the first/all occurrences of a given file content/commit?''. The \emph{compact}
data model, based on the novel notion of isochrone subgraphs, provides a
time/space trade-off that allows to track software provenance at the scale of
\SWH on consumer hardware. 
In future work we intend to extend the compact data model to allow tracking
provenance at the granularity of individuals lines of code, and explore how
other types of original source code artifacts evolve over time. We also intend
to study the characteristics of provenance graphs as naturally-occurring,
evolving complex networks.

\subsection{Notice}

We strongly believe that it is essential to make available to other researchers
the data on which is based this kind of analysis. The core of the work presented
here has been performed between January 2017 and August 2018, but the
unprecedented size of this dataset has required significant time and effort to
comply with our principles, and this has delayed the disclosure of our work much
longer than what we wanted or expected. Now that the \SWH Graph Dataset is
available to all~\cite{msr-2019-swh-dataset}, we are finally able to share
results that can be independently verified by other researchers.

\end{document}